\documentclass[10pt,a4paper]{article}
\newcommand{\equ}[1]{~Eq.~(\ref{#1})}
\newcommand{\Tr}{{\rm Tr\,}}

\newcommand{\be}[1]{\begin{equation}\label{#1}}
\newcommand{\ee}{\end{equation}}
\newcommand{\bea}[1]{\begin{eqnarray}\label{#1}}
\newcommand{\eea}{\end{eqnarray}}

\newcommand{\lless}{{\,<\!\!<\,}}

\newcommand{\half}{{\scriptsize \frac{1}{2}}}

\newcommand{\cas}{{\cal E}_{\rm Cas}}

\begin{document}

\title{Semiclassical Casimir Energies at Finite
Temperature}
\author{Martin Schaden$^{1,2}$ and Larry Spruch$^1$\protect\\
$^1$New York University, Physics Department \protect\\
4 Washington Place, New York, New York 10003\protect\\
$^2$Union College$^*$, Physics Department \protect\\
Science and Engineering Bldg., Schenectady, New York, 12308}
\date{\today}
\maketitle

\begin{abstract}
\noindent We study the dependence on the temperature $T$ of
Casimir effects for a range of systems, and, in particular, for a
pair of ideal parallel conductors, $l_1$ by $l_2$, separated by a
vacuum, a distance $l_3$ apart, with $l_1\gg l_3$ and $l_2\gg
l_3$. We study the Helmholtz free energy $A^T$, combining
Matsubara's formalism, in which the temperature $T$ appears as a
periodic Euclidean fourth dimension of circumference $l_T=\hbar
c/k_B T$ with the semiclassical periodic orbital approximation of
Gutzwiller. The latter was shown to be exact for parallel plates
at $T=0$. By inspecting the known results for the Casimir energy
at $T=0$ in two cases of a rectangular parallelepiped, ($l_1\gg
l_3$ and $l_2\gg l_3$, and $l_1\gg l_2$ and $l_1\gg l_3$), one is
led to guess at the expression for $A^T$ of two ideal parallel
conductors, without performing any calculation. The result is a
new form for $A^T$, namely, $A^T=-(2\hbar c l_1 l_2 l_3/\pi^2)
\sum_{n_3=1}^\infty\sum_{n_T=-\infty}^\infty L^{-4}(n_3, n_T)$,
where $L(n_3, n_T)=[(2 n_3 l_3)^2+ (n_T l_T)^2]^{1/2}$ is the
length of a classical periodic path on a two-dimensional cylinder 
section.
This expression for the free energy is equivalent to others that
have been obtained in the literature. At $T=0$ the semiclassical
approach provides a finite and systematic approximation scheme in
terms of classical paths that is useful when the normal modes of
the cavity cannot be determined either explicitly or implicitly.
Slightly extending the domain of applicability of Gutzwiller's
semiclassical periodic orbit approach, we here evaluate the free
energy  at $T>0$ in terms of periodic classical paths in a
four-dimensional cavity that is the tensor product of the original
cavity and a circle. The validity of this approach is at present
restricted to particular systems. We also discuss the origin of
the classical ($\hbar$-independent) form of $A^T$ for
$T\sim\infty$.
\end{abstract}

\vfill\noindent\rule[3pt]{1in}{1pt}{\small
\par\noindent $^*$Present address. Email address:
schadenm@union.edu}

\section{INTRODUCTION}
The best known of the Casimir effects is the force per unit area
between two parallel ideal plates, separated by vacuum, a distance
$l_3$ apart, at a temperature $T=0$\cite{a}. Some fifty years
later, that and other Casimir effects continue to intrigue both
theoreticians and experimentalists. Indeed, interest has been
growing. We list a few general references\cite{b,c,d,e}. Of
particular interest to us here are recent experimental reports by
Lamoreaux\cite{f} and by Mohideen and colleagues\cite{g}. Very
recently the possible role of this effect in
microelectromechanical systems has been examined\cite{g1}. To
avoid alignment problems one studies the force between a sphere of
radius $R$ and a wall, where the point on the sphere closest to
the wall is at a distance $l$ from the wall, with $l\lless R $.
(The analysis of the force in this case is intimately connected to
the force between walls\cite{b,c,h} and the experiment is simpler
to perform. An experiment which might be even simpler to perform,
because the force would be greater, is the force between a sphere
and a segment of a spherical shell. We are here dealing with the
focusing of virtual photons\cite{i}. Focusing of virtual photons
by a parabolic mirror has also been considered\cite{j}.) For the
values of $R$ and $l$ studied, the accuracy was such that a number
of corrections to the case of ideal conductors must be considered.
In particular, finite temperature effects must be accounted for.

The corrections to the Casimir force for $T>0$ were first studied
in a seminal paper by Lifshitz\cite{k} in 1956. The paper studied
dielectric slabs, and the temperature effects were a matter of
dispute for years. The subject was first clarified in
1978\cite{l}. It was shown that Lifshitz's basic results for the
temperature corrections were correct, but that Lifshitz had erred
in taking the limit as the dielectric constant goes to infinity
when going from a dielectric to an (ideal) conductor.

Our primary interest will be in rederiving the temperature effect
for ideal walls but in a new form. We will utilize two sets of
results, one by Matsubara\cite{m} for accounting for temperature
effects in a general setting, and the second a formulation of
Casimir effects at $T=0$ in terms of semiclassical extremal paths
by the present authors\cite{n}. The reformulation has the merit
that the dominant paths are often rather obvious, with
contributions that can be evaluated even when the proper frequency
modes cannot be determined.

We take this opportunity to mention two significant papers\cite{o}
by Balian and Duplantier not referred to in\cite{n}. The methods
used in\cite{o} were quite different from those used in our paper.
We had not fully appreciated the two papers at the time\cite{n}
was published.

\section{IDEAL PARALLEL PLATES: A HEURISTIC APPROACH}
Matsubara showed\cite{m} that in studying a system in
three-dimensional space in equilibrium at temperature $T$, it can
be convenient to consider the system in a four-dimensional
Euclidean space whose fourth dimension at finite temperature
$T>0$ is compactified to a circle of circumference \be{circ}
l_T=\hbar c/(k_B T)\ . \ee Thus, consider two ideal parallel
conductors at temperature $T$, with an area ${\cal A}=l_1 l_2$
separated a distance $l_3$ in vacuum with $l_3\lless l_1$ and
$l_3\lless l_2$. The force between the plates in this case is
proportional to $l_1$ and $l_2$ and the force per unit area does
not depend on either dimension. At finite temperature, the
essential geometry in this case is that of a two-dimensional
cylindrical surface of length $l_3$ and circumference $l_T$. (See
App.~A)

In our study\cite{n} of the plates at $T=0$ the analysis
involved the determination of the {\it difference}, $\Delta \rho(E)$, of the
spectral densities in the presence and absence of the conductors.
Gutzwiller\cite{p} showed that this quantity is given
in semi-classical approximation by the {\it periodic} paths of extremal {\it classical} action,
in our case the periodic paths of extremal length. (Note that none of these
paths has zero length and that a {\it stationary} periodic path must
involve a minimum of two reflections.) In the case of parallel
plates these are classical paths going back and forth between, and
perpendicular to, the walls.

The analogous analysis at finite temperature would involve studying the
semiclassical Green function with a periodic dependence on the Euclidean
time. The initial and final space-time points are still the same, but the
Euclidean geometry is that of a cylinder. We prefer to proceed differently.

For $T>0$ (and fixed volume),  it is natural to work with the
Helmholtz free energy $A^T$ rather than with ${\cal E}^T$, the
(Casimir) energy itself, and there is a natural choice of the
form of $A^T$. Thus, abandoning any pretense of rigor, we guess
at the form of $A^T$ -- there remains the determination of an
over-all numerical factor -- and justify our choice of $A^T$
afterwards.

The semi-classical path approach has advantages that go well
beyond providing a simple physical picture of the origin of the
Casimir effect at $T>0$; as noted earlier, one can often determine
the significant paths even when the natural frequency modes cannot
be determined (neither explicitly nor, as in the case of the
generalized argument theorem\cite{Kampen,q}, implicitly).

We have \be{free} A^T=A^T(\hbar,c,l_1,l_2,l_3,l_T). \ee At $T=0$,
we were concerned with the energy density. At $T>0$, we will be
concerned with the free energy density. In both cases we
explicitly extract the volume $l_3 {\cal A}$ of the cavity.
Further, on dimensional grounds, $A^T$ is proportional to $\hbar
c$. We can therefore write \be{factor} A^T=\hbar c l_3 {\cal A}
g(l_3,l_T)\ ,
\ee where $g$ is proportional to the inverse fourth power of a
length. Now, in\cite{n} it was shown that for a cavity with
$l_3\lless l_1,l_2$, ${\cal E}_{\rm Cas}$ is proportional to
$\sum_{n_3} L^{-4}(n_3)$, where the length $L(n_3)=2n_3 l_3$ and
$n_3=\pm 1,\pm 2,\dots$. Further, for a parallelepiped with
$l_3\lless l_1$ and $l_2\lless l_1$, ${\cal E}_{\rm Cas}$ was
proportional to $\sum_{n_2}\sum_{n_3} L^{-4}(n_2,n_3)$, where
$L(n_2,n_3)= [(2n_2 l_2)^2+(2n_3 l_3)^2]^{1/2}$ and the sum is
over all pairs of integers $(n_2,n_3)$, positive, negative, and
zero, other than $(0,0)$. $L(n_3)$ is the length of a classical
periodic path perpendicular to the walls and reflected $|n_3|$
times from each of them. Positive values of $n_3$ refer to paths
which start to the right, while negative values of $n_3$ refer to
paths which start to the left. $L(n_2,n_3)$ is the length of a
classical periodic path with $|n_2|$ reflections off each of the
walls with dimensions $l_1$ by $l_3$ and $|n_3|$ reflections off
each of the walls with dimensions $l_1$ by $l_2$. For the case
with $l_3\lless l_1$ and $l_3\lless l_2$, $T>0$, to which we
henceforth largely restrict our attention, the knowledge that the
space is enlarged by one extra Euclidean dimension strongly
suggests that $A^T$ can be obtained by simply replacing
$L(n_2,n_3)$ by \be{length} L(n_3,n_T)=[(2n_3 l_3)^2+(n_T
l_T)^2]^{1/2} \ee This length is that of a classical periodic
extremal path on the surface of a cylinder of height $l_3$ and
circumference $l_T$. The path starts in a given direction and
reflects $|n_3|$ times from each of the ends of the cylinder
section (the ideal walls) and  circles the cylinder in a
particular direction $|n_T|$ times. (The absence of a factor of 2
in the $n_T l_T$ term of\equ{length} is due to the fact that we
are considering extremal  paths on a section of a
 {\it cylindrical} surface. See App.~A)  Negative values of $n_3$
refer to paths that start in the opposite direction, while negative
values of $n_T$ refer to paths that
circle the cylinder in the opposite direction. The contribution from
the $n_3=0$ path can be ignored since it gives an (extensive)
 contribution to the free energy that is proportional to the volume; it
does  not contribute to the (net) force on a plate. [See paragraph
in Sec.~4 above the paragraph which contains\equ{boxclass}.]

We thus guess that 
\be{finc} A^T=K\hbar c {\cal A} l_3\sum_{(n_3,
n_T)} L^{-4}(n_3,n_T)\ , \ee where the sum extends over all pairs
of integers $(n_3,n_T)$ with $n_3\neq 0$, the length $L(n_3,n_T)$
is given by\equ{length}, and $K$ is a numerical coefficient. We
determine the value of $K$ by demanding that $A^T$ reduce to
\be{cas0} {\cal E}_{\rm Cas}=-\frac{\pi^2\hbar c {\cal A}}{720
l_3^3} \ee for $T\sim 0$. But $T\sim 0$ implies that
$l_T\sim\infty$ which in turn implies that only $n_T=0$
contributes to the sum in\equ{finc}. For $T\sim 0$ we thus have
\be{fin0} A^{T\sim 0}=K\hbar c {\cal A} l_3
2\sum_{n_3=1}^{\infty}\frac{1}{(2 n_3 l_3)^4}=\frac{\pi^4 K\hbar
c {\cal A}}{720 l_3^3}\ .
\ee
(The factor of 2 reflects the fact
that we are now summing over positive values only of $n_3$.)
Comparison with\equ{cas0} gives $K=-1/\pi^2$ and we obtain
\be{fin} A^T=-\frac{2\hbar c {\cal A}
l_3}{\pi^2}\sum_{n_3=1}^\infty \sum_{n_T=-\infty}^\infty
L^{-4}(n_3,n_T)\ .
\ee
In App.~B we show that\equ{fin} is indeed correct by comparing to
a previous result. A proof from first principles, for a scalar
field, is given in Sec.~5. The force per unit area at a
temperature $T$ is then given by \be{forceT} \frac{F^T}{{\cal
A}}=\frac{1}{{\cal A}}\frac{\partial}{\partial l_3}A^T =
-\frac{2\hbar c}{\pi^2}\sum_{n_3=1}^\infty
\sum_{n_T=-\infty}^\infty \frac{3 (2 n_3 l_3)^2-(n_T
l_T)^2}{[(n_T l_T)^2+(2 n_3 l_3)^2]^3} \ee The somewhat
complicated form of $F^T$ as opposed to the relatively simple
form of $A^T$ shows that the free energy is not just a more
fundamental object of study but also a more transparent one.
Given $A^T$, $F^T$ follows as in\equ{forceT}.

\section{THE HIGH-T AND LOW-T EXPANSIONS}
The free energy $A^T$  is readily obtained from\equ{fin} for $z$
large  or small, where
\be{z}
z=2 l_3/l_T=2 l_3 k_B T/(\hbar c)\ .
\ee
Isolating  the $n_T=0$ contribution, the free energy of\equ{fin}
may be rewritten in the form
\be{defdelta}
\frac{A^T(l_3)}{{\cal A}}=-\frac{\pi^2\hbar c}{720 l_3^3}-
\frac{ (k_B T)^2}{\pi^2\hbar c l_3}\Delta(z)\ .
\ee
 {From} \equ{fin} and the
definition\equ{length} we have that the dimensionless function
$\Delta(z)$ is
\be{refld} \Delta(z)=\sum_{n=1}^\infty
\sum_{m=1}^\infty [m^2 z+n^2/z]^{-2}=\Delta(1/z).
\ee
[To check\equ{refld}, let $z\rightarrow 1/z$, and interchange $m$
and $n$.] The reflection property\cite{BM},
$\Delta(z)=\Delta(1/z)$, is quite powerful. It relates the high-
and low- temperature regimes, or, equivalently, the large and
small $l_3$ regimes. The expansion of $\Delta(z)$ for $z\gg 1$ is
found by using the identity\cite{MF}
\be{exp1} \sum_{n=1}^{\infty}\frac{1}{n^2+a^2}\equiv
J(a)/a=\frac{1}{2 a^2}(a\pi \coth(a\pi)-1)=-\frac{1}{2
a^2}+\frac{\pi}{2 a}+\frac{\pi}{a} \sum_{n=1}^\infty e^{-2\pi n
a}\ . \ee Differentiating\equ{exp1} with respect to $a^2$ and then
replacing $a$ by $m z$ one obtains, after multiplication by $z^2$
and summation over $m$, \bea{exp2} \Delta(z)&=&\sum_{m=1}^\infty
\left[-\frac{1}{2 m^4 z^2}+\frac{\pi}{4 m^3 z} +\frac{\pi}{2 m^3
z}\sum_{n=1}^\infty (1+2 \pi m n z)e^{-2 \pi m
n z}\right]\nonumber\\
&=&-\frac{\pi^4}{180 z^2}+\frac{\pi}{4 z}\zeta(3)+\frac{\pi}{2
z}\sum_{n=1}^\infty e^{-2\pi n z} (1+2\pi n
z)\!\!\!\!\!\!\!\sum_{n/m {\rm~integer}>0}{\hskip-2em} m^{-3}\ ,
\eea
where $\zeta(3)=\sum_{m=1}^\infty m^{-3}=1.202\dots$. Using the
expansion of\equ{exp2} in\equ{defdelta} gives for the free energy
per unit area, for $2 l_3 k_B T>\hbar c$, or $2 l_3>l_T$,
\be{fasympg}
\frac{A^T}{{\cal A}}=-\frac{\zeta(3) k_B T}{8\pi l_3^2}-\frac{(k_B
T)}{l_T l_3} (1+\frac{l_T}{4\pi l_3}) e^{-4 \pi l_3/l_T}+
O\left(\frac{k_B T}{l_3^2} e^{-8\pi l_3/l_T}\right)\ .
\ee
Using the expansion of\equ{exp2} with $z$ replaced by $1/z$ in\equ{defdelta}
gives, for $2 l_3 k_B T<\hbar c$, or $2 l_3<l_T$,
\be{fasympl}
\frac{A^T}{{\cal A}}=-\frac{\pi^2 \hbar c}{720 l_3^3}-\frac{k_B T}{2\pi
l_T^2}\zeta(3)+\frac{\pi^2 k_B T l_3}{45 l_T^3}-\frac{k_B
T}{l_T l_3}(1+\frac{l_3}{\pi l_T}) e^{-\pi l_T/l_3} + O\left(\frac{k_B T}{l_T l_3} e^{-2\pi l_T/l_3}\right)\ .
\ee

These low-T (\equ{fasympl}) and high-T (\equ{fasympg}) expansions
of $A^T/{\cal A}$, first obtained by Sauer, by Mehra, and by Levin
and Rytov\cite{SM}, have also been reproduced in\cite{l}. (The
present derivation is somewhat simpler, to some extent because
some of the earlier derivations were in a broader context, with
dielectric walls or rather arbitrary surfaces. Note that
Mehra\cite{SM} uses a convention for which an {\it attractive}
force between the plates is taken to be positive.) See\cite{c} for
a physical explanation of the contribution proportional to $T^4$
in the low-T expansion\equ{fasympl} of the free energy.

The expansions\equ{fasympl} and\equ{fasympg} are both
exponentially good and the two limits are ``dual'' in the sense
that since $\Delta(z)=\Delta(1/z)$, they {\it both} follow from
\equ{exp2}. The relative error of these approximations is largest
at the transition point, $2 l_3 k_B T= \hbar c$, or
$z=2l_3/l_T=1$. (The $l_3$ dependence of $A^T/{\cal{A}}$ is not
well defined at the transition point, since, with $n$ arbitrary, a
factor of $(2l_3/l_T)^n\sim 1$ does not change the function
appreciably near this point.) The exponential corrections all have
the same sign and are no greater than $\sim 2.3\%$ at any value of
$l_3 k_B T$. Perhaps most significant experimentally is that the
high-temperature limit is valid for $l_3$ rather large compared to
$\hbar c/(2 k_B T)$. This implies that the fall-off of the Casimir
force at room temperature is proportional to $l_3^{-3}$ rather
than $l_3^{-4}$ for separations  $l_3$ rather large compared to $
4 {\rm ~microns}$. Since the largest separation between the plates
of early experiments was a few microns, it appears unlikely that a
Casimir force proportional to $1/l^4$ could have been observed
over the entire range. It should, however, be recalled that we
have here considered an ideal metal rather than a real one. On the
other hand, corrections due to finite conductivity {\it diminish}
with increasing separation and are expected to be negligible
compared to the temperature correction for separations exceeding a
few microns\cite{c,SL}.

\section{THE CLASSICAL FORM OF $A^T$ FOR $T\rightarrow\infty$}
The fact that for parallel plates $A^T$ at sufficiently large $T$
assumes a classical ($\hbar$-independent) form, as seen
in\equ{fasympg}, warrants some comment. We begin with the remark
that $A^T$ in the high-T limit is not generally
$\hbar$-independent. Thus, for a thin smooth closed surface, one
finds\cite{o} \be{smooth} A^T=-\sigma k_B T \ln(k_B T l/\hbar c)\
, \ee where the dimensionless quantity $\sigma$ is a function of
{\it ratios} of the radii of curvature and $l$ is an
appropriately defined overall length scale. The logarithmic
dependence of \equ{smooth} on the overall length scale and $k_B
T/\hbar c$ implies that the  {\it change} of the free-energy does
not depend on $\hbar c$ when all lengths are rescaled by the same
amount, i.e., the ``breathing mode'' is always classical at high
temperatures.

The fact that the free energy is a pure power law for parallel
plates makes that case a bit special. If we {\it assume} that the
high-T limit of the free energy of two parallel plates is
classical, the power law dependence of $A^T$ follows on
dimensional grounds. (It follows from $A^T=A^T(k_B T,l_1,l_2,l_3)$
that $A^T=k_B T l_1 l_2 g(l_3)$, where $g(l_3)$ must be
proportional to $1/l_3^2$.) One may argue for the existence of a
classical high-T limit (and the absence of logarithmic
corrections)  from the contribution to the
free energy of a single cavity mode of frequency $\omega_k$.
Writing 
\be{defsingle} 
A^T=-\beta^{-1} \ln Z=\sum_k A_k^T
=\sum_k\left[\frac{\hbar\omega_k}{2}
+\beta^{-1}\ln(1-e^{-\beta\hbar \omega_k})\right]\ , \ee $A^T_k$
for frequencies $\omega_k\lless k_B T/\hbar$ becomes \be{single}
A^{T\sim\infty}_k=\frac{\hbar\omega_k}{2} +\beta^{-1}
\ln[\beta\hbar\omega_k (1-\frac{\beta\hbar\omega_k}{2}+
O((\beta\hbar\omega_k)^2))]\sim k_B T \ln(\frac{\hbar\omega_k}{k_B
T})\ , 
\ee 
ignoring terms of order $\beta (\hbar\omega_k)^2$.
Inserting the high-T expression\equ{single} in\equ{defsingle}
formally gives an asymptotic expansion of $A^T$. But since the
frequencies of a cavity generally can be arbitrarily high and the
neglected terms are not small for {\it all} frequencies at a fixed
value of $T$, this expansion is valid only in the presence of a
cutoff. However, for the {\it change} of the free energies of the
cavity in two different configurations the mode sum is finite if
the two configurations can be connected {\it adiabatically}, as,
for instance, by moving an internal boundary arbitrarily slowly.
An asymptotic expansion for large $T$ of this difference does make
sense. Under an adiabatic change of boundary conditions, the
eigenmodes of the cavity change from $\omega_k$ to
$\omega_k^\prime$, but the quantum numbers of a mode are
conserved. Using\equ{single} one obtains for the asymptotic
difference of the free energies when $T\sim\infty$
\be{diff}
\Delta A^T=A^{\prime T}-A^T\sim k_B T\sum_k\left[
\ln(\frac{\omega^\prime_k}{\omega_k})\right]\ .
\ee The sum in\equ{diff} converges if the difference of the free
energies of the two configurations has a finite limit at $T=0$,
since a finite temperature does not induce new ultraviolet
divergences. The expression\equ{diff} for the difference of the
free energies at sufficiently high temperature (or better, for
$k_B T l\gg \hbar c$, where $l$ is a relevant length scale of the
problem) is classical and notably does not depend on $\hbar$. In
this limit the entropy $S$ of the cavity is $S=-A^T/T$
and\equ{diff} states that the difference in the entropies of a
cavity at high temperatures approaches a constant that does not
depend on the quantum scale $\hbar$. (That the entropy of a
bosonic gas in a cavity has a {\it finite} limit as
$T\rightarrow\infty$ is compatible with dimensional reduction,
that is, the high temperature limit is described by Euclidean
quantum field theory in $3$ dimensions with $\hbar$ replaced by
$i\beta$\cite{m}.)

Although plausible, it is not possible to prove the existence of
a finite high temperature limit using {\it only} classical
arguments, because one can only compare the energy $k_B T$ with
the frequencies of the cavity by introducing a quantity with the
dimension of $\hbar$. If the asymptotic limit exists, and we have
just argued that it does, it is independent of the value of
$\hbar$ one has chosen to make this comparison. On dimensional
grounds, the classical contribution to the change in entropy from
a single (independent) mode can then depend only on the frequency
ratio. Further, since the contribution to the (change in) entropy
of (independent) modes is additive, classically the (change) in
entropy due to a single mode is proportional to
$\ln(\omega^\prime/\omega)$. \equ{diff} then follows as the
classical change in the free energy of a cavity from the
definition of $k_B$.

In studying the force between $l_1$ by $l_2$ plates at a
separation $l_3$ one can evaluate the energy in the volume
between the plates and in the infinite volume outside the plates
and then determine the force by differentiation with respect to
$l_3$. It is a standard practice to avoid infinite volumes by
considering a box of fixed volume\footnote{There should be no
confusion with our usage of $L$ to also denote path lengths.}
$l_1\times l_2\times L$ with an additional (movable) plate
parallel to the $l_1$ by $l_2$ walls at a distance $l_3$ from one
of the walls\cite{P}. The total free energy of the box is the sum
of the free energies of each of the two subvolumes:
$A^T_{tot}(l_1,l_2,L;l_3)=A^T(l_1, l_2, l_3)
+A^T(l_1,l_2,L-l_3)$. The force on the additional wall is
$-\partial A^T_{tot}/\partial l_3$. Extensive contributions to
the free energy of the subsystems (proportional to their volume)
do not contribute to the dependence of $A^T_{tot}$ on $l_3$ and
therefore do not result in a (net) force on the middle wall. Note
that generally only the {\it difference} of the free energy of
the box compared to the free energy of the box in some
``standard'' configuration, (say with the additional wall at
$l_3=L/2$) i.e., $A^T_{diff}(l_1,l_2,L;l_3)
=A^T_{tot}(l_1,l_2,L;l_3)-A^T_{tot}(l_1,l_2,L;L/2)$, is finite.

With these arguments, we can, purely classically, obtain the high
temperature limit of the change in the free energy, $\Delta A^T$,
of a scalar field in a box of dimension $l_1\times l_2\times L$,
 as we move an additional wall from a position at a distance
$l_3$ from one of the $l_1$ by $l_2$ sides to the middle of the
box at $L/2$. Thus, from\equ{diff}, we have
\be{boxclass}
\Delta A^T_{class}=k_B T {\cal A}\int \frac{d^2 k_\perp}{(2\pi)^2}
\half \sum_{n=-\infty}^\infty \left[\ln\frac{\omega(\pi n/l_3,
k_\perp)}{\omega(2 \pi n/L,k_\perp)}+\ln\frac{\omega(\pi n/[L-l_3],
k_\perp)}{\omega(2 \pi n/L,k_\perp)}\right]
\ee
where $\omega(k_3,k_\perp)=c\sqrt{k_3^2+k_\perp^2}$, with
$k_3=\pi n/\bar l_3$ where $\bar l_3$ is the appropriate distance
from the wall. Using the integral representation
\be{lnrep}
\ln[a/b]=\half\int_0^\infty \frac{d\lambda}{\lambda}[e^{-\lambda b^2}
-e^{-\lambda a^2}]\ ,
\ee
for the logarithms in\equ{boxclass}, the integration over
$k_\perp$ can readily be performed on using polar coordinates. In
addition we use the reflection property,
\be{refl}
\sum_{n=-\infty}^\infty e^{-n^2 \pi^2 x}=(\pi
x)^{-1/2}\sum_{n=-\infty}^\infty e^{-n^2/x}\ ,
\ee
to rewrite the resulting sums over $n$. [\equ{refl} and
\equ{refld} both relate the values of a function at the argument
$x$ and at the "reflected" argument $1/x$. \equ{refl} is a
particular case of Poisson's summation formula.] We then obtain
\be{boxclass1}
\frac{\Delta A^T_{class}}{k_B T {\cal A}}=-\frac{1}{16\pi
\sqrt{\pi}} \sum_{n=-\infty}^\infty \int_0^\infty
\frac{d\lambda}{\lambda^{5/2}}[l_3 e^{-\frac{n^2 l_3^2}{\lambda}}
+ (L-l_3) e^{-\frac{n^2 (L-l_3)^2}{\lambda}}- L e^{-\frac{n^2
(L/2)^2}{\lambda}}]\ .
\ee
The $n=0$ contribution vanishes and $\sum_{n=-\infty}^\infty$
reduces to $2 \sum_{n=1}^\infty$. The integrals with $n\neq 0$
converge term by term; with
\be{integrals}
\int_0^\infty\frac{d\lambda}{\lambda^{5/2}}
e^{-\alpha/\lambda}=\frac{\pi^{1/2}}{2\alpha^{3/2}}\ ,
\ee we thereby arrive at
\be{classres} \Delta A^T_{class}=-k_B T
{\cal A}\frac{1}{16\pi}\zeta(3)
[\frac{1}{l_3^2}+\frac{1}{(L-l_3)^2}-\frac{8}{L^2}]\ ,
\ee
which for $L\gg l_3$ is the leading term of\equ{fasympg} up to a
factor of $2$; the factor appears because we considered the scalar
case here.[We strongly suspect that there may be a simpler {\it
classical} derivation of\equ{classres} than the one given here.]

A related problem is that of the Casimir energy, at $T=0$, of two
spinless charged particles in free space. There is a classical
contribution in that case too, but it is {\it not} the leading
term and the question arises if the term can be evaluated using
classical theory. The answer in this case is also
positive\cite{Sucher}.

An interesting point is that on dimensional grounds the absence of
$\hbar$ leads to the absence of $c$, though we are studying
electromagnetic waves. There is no contradiction; the classical
Maxwell equations have meaning in the non-relativistic
($c\rightarrow\infty$) limit\cite{SS}. Thus, for example, in the
dipole approximation the classical Maxwell equations give a
$c$-independent radiative transition rate.

\section{THE SCALAR FIELD}

In App.~B we prove the validity of\equ{fin} by making a
connection with the results of\cite{l}. We here give an ab initio
proof of that equation for a massless scalar field. The relation
to the electromagnetic case is discussed in App.~C (in a less
broad context than that given, for example, in ref.\cite{o}).

Thus, to substantiate some of the arguments of the main text, we
consider a free massless scalar field $\phi$ that satisfies
Dirichlet boundary conditions
\be{bplates}
\phi(x_i=0)=\phi(x_i=l_i)=0\ ,
\ee
at the boundaries of a box with spatial dimensions $l_1\times
l_2\times \dots \times l_D$. The box serves as an infrared cutoff
and guaranties a discrete spectrum. The dimension $D$ of the box
is arbitrary and can be used to dimensionally regularize the
ultraviolet behavior of subsequent expressions by analytic
continuation to non-integer dimensions. Additional boundary
conditions may be imposed on the field inside the box. One could,
for instance, demand that the field also vanish on a  spherical
shell located within the box, or, more appropriately for the
Casimir force between plates, that it also vanish on a plane
within the box.

The free energy $A^T$ is related to the partition function $Z$ by
$A^T=-\beta^{-1}\ln Z$, where  $\beta=1/(k_B T)$. Formally, $Z$ is
\be{partition}
Z=\Tr e^{-\beta H}\ ,
\ee
where $H=\sum_{\bf k} E_{\bf k} (a^\dagger_{\bf k} a_{\bf k}+1/2)$
is the free hamiltonian. $a_{\bf k}$ ($a^\dagger_{\bf k}$)
destroys (creates) a quantum with quantum numbers ${\bf k}$ in the
(possibly rather complicated) cavity. Crucial for the following is
that $H$ [ignoring any (weak) self-coupling of the scalar field]
commutes with the number operators $n_{\bf k}=a^\dagger_{\bf k}
a_{\bf k}$ no matter how complicated the boundary conditions on
the field may be. $H$ can therefore be diagonalized in a basis
that enumerates the number of quanta in each cavity mode and has
the eigenvalues $\sum_{\bf k} E_{\bf k}(n_{\bf k}+1/2)$ with each
$n_{\bf k}$ a positive integer or $0$. Evaluating the trace
of\equ{partition} in this number basis, one has
\be{ltr}
\ln Z=\ln[\prod_{\bf k}\sum_{n_{\bf k}=0}^\infty e^{-\beta
(n_{\bf k}+1/2) E_{\bf k}}]=-\sum_{\bf k}\ln[2\sinh(\beta E_{\bf
k}/2)]\ .
\ee
Instead of computing the free energy or partition function
directly, it will be convenient to first consider the energy at
temperature $T$,
\be{enerdef}
{\cal E}^T=\frac{\partial}{\partial \beta} (\beta
A^T)=-\frac{\partial}{\partial \beta} \ln Z\ .
\ee
Differentiating\equ{ltr}, one finds
\be{u}
{\cal E}^T=\sum_{\bf k} E_{\bf k} [\frac{1}{2}+\sum_{m=1}^\infty
e^{-m\beta E_{\bf k}}]\ .
\ee
For $E_{\bf k}>0$ one can use Cauchy's theorem with a contour
that runs just above the real axis and is closed by a large
semicircle in the upper complex plane and write
\be{Ct}
E_{\bf k}[\frac{1}{2}+\sum_{m=1}^\infty e^{-m\beta  E_{\bf
k}}]=-\lim_{\epsilon\rightarrow
0^+}\frac{1}{\pi}\int_{-\infty+i\epsilon
}^{\infty+i\epsilon}\frac{\lambda^2 d\lambda}{E_{\bf
k}^2+\lambda^2}[1/2+\sum_{m=1}^{\infty} e^{im\beta \lambda}]\ ,
\ee
up to terms that do not depend on $E_{\bf k}$ nor on $\beta$;
in\equ{Ct} we have dropped the contribution from the semicircular
part of the contour. (For a sufficiently large radius $\Omega$ of
the contour, this contribution is $\Omega/\pi$ up to terms that
vanish for $\Omega\rightarrow \infty$.) The integral along the
semicircular part of the contour thus does not depend on $E_{\bf
k}$ nor $T$ and can be absorbed in the overall normalization of
the energy. Such constant contributions are irrelevant for the
discussion of energy differences. With $\lambda=x+i\epsilon$ and
$x$ real, the term on the right hand side of\equ{Ct}, in square
brackets, is\cite{sumis}
\be{sumr}
Q\equiv[1/2+\sum_{m=1}^{\infty} e^{im\beta
(x+i\epsilon)}]=\frac{i}{2} \cot(\beta x/2)+\pi
\sum_{n=-\infty}^\infty \delta(\beta x-2\pi n) +O(\epsilon)\ .
\ee
We can now evaluate the integral in\equ{Ct}. Apart from terms of
order $\epsilon$ that vanish in the limit $\epsilon\rightarrow
0^+$, we can replace $\lambda$ by $x$. Because $\cot(\beta x/2)$
is an odd function of $x$, the imaginary part of $Q$ does not
contribute to the integral. (This reflects the assumption that the
cavity is ideal.) Formally, the energy of an ideal cavity at
finite temperature is thus, from Eqs.~(\ref{u}), (\ref{Ct}), and
(\ref{sumr}),
\bea{fr1}
{\cal E}^T &=&-\sum_{\bf k}\sum_{n=-\infty}^\infty \frac{(2\pi
n)^2}{\beta^3} E^{-2}({\bf k},n)=-\frac{\partial}{\partial\beta}
\sum_{\bf k}\sum_{n=-\infty}^\infty\, \ln E({\bf k
},n)\nonumber\\&=&-\frac{\partial}{\partial\beta}
\ln\left[\prod_{\bf k}\prod_{n=-\infty}^\infty\, E^{-1}({\bf
k},n)\right]\ ,
\eea
where
\be{Ek}
E({\bf k}, n)=[E^2_{\bf k} + (2\pi n/\beta)^2]^{1/2}\ .
\ee
Since ${\cal E}^T=-(\partial/\partial\beta) \ln Z$, \equ{fr1}
determines the partition function up to an overall normalization,
or equivalently, the entropy up to an overall constant (which can
be determined by demanding that the entropy vanish at $T=0$).
\equ{fr1} formally shows that the partition function is
proportional to the square root of the inverse determinant of the
differential operator $H_4^2=h^2-(\hbar c)^2
\frac{\partial^2}{\partial x_T^2}$, where $x_T$ is the additional
Euclidean coordinate in which the fields are periodic (with
periodic length $l_T$) and $h$ is the differential operator
(single particle hamiltonian) with eigenvalues $E_{\bf k}$. The
positive eigenvalues of $H_4$, the $E({\bf k},n)$ given
in\equ{Ek}, are the allowed energies of a massless scalar in the
{\it four}-dimensional cavity constructed by augmenting every
point of the original three-dimensional cavity by a circle of
circumference $l_T$, that is, the 4-dimensional space is the
tensor product of the cavity and a circle whose circumference is
proportional to the inverse temperature. From\equ{fr1} and
\equ{enerdef} we see that the free energy of the cavity can be
written formally as
\be{AT}
A^T_{\rm cavity}=\beta^{-1}\sum_{\bf k}\sum_{n=-\infty}^\infty\,
\ln E({\bf k},n)=\frac{\hbar c}{l_T} \int_0^\infty dE^\prime \rho^T_{\rm
cavity}(E^\prime) \ln(E^\prime)\ ,
\ee
with
\be{defrho}
\rho^T_{\rm cavity}(E^\prime)=\sum_{\bf
k}\sum_{n=-\infty}^\infty\, \delta(E({\bf k},n)-E^\prime)\ .
\ee
\equ{defrho} shows that $\rho^T_{\rm cavity}(E^\prime)$  is the
spectral density of the {\it four}-dimensional cavity constructed
from the three-dimensional one in the above manner. Changes of
$\rho^T_{\rm cavity}(E^\prime)$ due to (adiabatic) deformation of
the cavity, including a change of temperature, can then be
approximated semi-classically using periodic classical rays of the
{\it four}-dimensional cavity. This leads to a description of the
free energy in terms of periodic paths in the {\it
four}-dimensional cavity which we exploited when we guessed at the
form of the free energy in Sec.~2.

To be somewhat more concrete, consider as an  example 
a cubic cavity with sides of length $\bar l$. The momentum components 
for the mode with energy $E_{\bf k}$ are $k_i \hbar=n_i\pi \hbar/\bar
l$, and the energies in the corresponding
four-dimensional cavity at finite temperature are 
\be{ex5}
E({\bf k}, n)=\pi\hbar c\left[\sum_{i=1}^3  \left(\frac{n_i}{\bar
l}\right)^2 + \left(\frac{n}{l_T/2}\right)^2\right]  \ ;
\ee
as always, the length associated with the fourth (periodic) dimension,
the analog of the lengths associated with the spatial dimensions, is
$l_T/2$.

Note that\equ{AT} is formal in two respects: the integral over the
energy diverges due to the ultraviolet behavior of the spectral
density, and \equ{AT} specifies the free energy only up to (a
generally similarly divergent) contribution proportional to $T$.
Considerably more rigorous and physically relevant expressions can
be obtained from\equ{AT} for {\it changes} in the free energy due
to adiabatic deformations of the three-dimensional cavity. In
physical terms the difference in the free energy is work that has
to be done to change the boundary conditions for the field and
should therefore be finite whenever it is physically feasible to
do so. The ambiguity that\equ{AT} defines the entropy only up to a
constant that does not depend on the temperature is removed by
requiring that the entropy vanish at $T=0$.

Let us finally remark that for $T\rightarrow 0$ and $E_{\bf k}$
fixed,
\be{lim0}
\lim_{\beta\rightarrow\infty}\sum_{n=-\infty}^{\infty}
\beta^{-1}\ln[E_{\bf k}^2+(2\pi n/\beta)^2]^{1/2}= E_{\bf k}/2,
\ee
up to an infinite constant that does not depend on $E_{\bf k}$ and
therefore cancels in differences. [\equ{lim0} follows upon
subtracting the $E_{\bf k}$-independent term $\ln[\mu^2 +(2\pi
n/\beta)^2]^{1/2}$ from the summand in\equ{lim0} and noting that
the so regularized  sum defines a convergent integral in the limit
$\beta\rightarrow 0$ with the value $(E_{\bf k}-\mu)/2$. The
dependence on the subtraction point $\mu$ can be absorbed in the
overall energy normalization and cancels in energy differences.]
At $T=0$ \equ{AT} thus gives the usual expression for the Casimir
energy. The classical periodic orbits of the four-dimensional
cavity in this limit are just the periodic orbits of the
three-dimensional cavity, since any path that winds about the
fourth dimension becomes arbitrarily long as $T\rightarrow 0$.
Note that whereas the spectral density $\rho_{\rm cavity}^T$
generally is proportional to the four-dimensional volume $l_T
V_{\rm cavity}$ the factor $1/l_T$ in\equ{AT} ensures
that $A^T$ generally will only be proportional to the
three-dimensional volume of the cavity (which we extracted
explicitly in Sec.~2).

Although our starting point, \equ{fr1} for the free energy, is
the field theoretic one\cite{m}, the evaluation of this
expression using the spectral density of a massless scalar in four
spatial dimensions and, further, the semiclassical approximation
to (changes in) this spectral density using classical periodic
rays of the four-dimensional cavity is perhaps a bit unusual. This
sketch of the relation between the free energy and the periodic
classical rays of a corresponding four-dimensional cavity is the
mathematical and perhaps even the physical basis for our guess of
the free energy of two parallel plates in Sec.~2.

\section{DISCUSSION}
For ideal parallel plates, the semiclassical periodic orbital
formalism gives a result for $\cas$ different in form but
equivalent to that obtained by other methods. However, the
approach provides a simpler physical picture, allows a simpler
transition from $\cas$ to $A^T$, and gives a form simpler to
evaluate numerically. It evaluates changes in the free energy due
to changes in the temperature and in the cavity boundaries using
the periodic classical paths in a four-dimensional cavity which is
the tensor product of the original three-dimensional one and a
circle. In a number of instances this amounts to replacing the
contribution to the Casimir energy of each periodic classical
path of length $L_\gamma$ at $T=0$ by the sum of contributions of
periodic classical paths of length
$L_\gamma^T(n_T)=[L_\gamma^2+(n_Tl_T)^2]^{1/2}$ that, in
addition, wind about the extra fourth dimension $n_T$ times.

Furthermore, preliminary research indicates that this considerable
simplification may  be present for a number of geometries, for
scalar as well as for electromagnetic fields, and, at least for
parallel plates, for dielectric walls. The simplification is a
slight extension of the domain of applicability of Gutzwiller's
semi-classical periodic orbit approach, allowing the  method to be
used for a number of systems at finite temperature. Generalizing
the result for parallel plates, it appears that the periodic
classical paths of various lengths can be interpreted as dual
variables to the eigenfrequencies of a cavity: in rectangular
cavities the two descriptions are related by Poisson's summation
theorem.

\begin{center}
{\bf ACKNOWLEDGEMENTS} \end{center} \noindent This work was
supported by the National Science Foundation Grant PHY-0070525.
M.S. greatly enjoyed the hospitality of New York University where
most of this work was done.
\appendix
\setcounter{equation}{0}
\renewcommand\theequation{\Alph{section}\arabic{equation}}
\renewcommand\thesection{APPENDIX~\Alph{section}:~}
\section{FINITE TEMPERATURE AND THE INTRODUCTION OF A FOURTH DIMENSION}
As a vague indication of why the temperature can be treated by
introducing a periodic fourth dimension, compare the standard
operators $\exp(-iHt/\hbar)$ in quantum theory and $\exp(-\beta
H)$ in statistical mechanics. One finds $\beta=it/\hbar$. The
presence of $i$ suggests periodicity, and we therefore drop the
factor $i$ and replace $t$ by a period $\tau$, giving
$\beta=\tau/\hbar$. For photons, we can replace $\tau$ by $l_T/c$,
leading to $l_T=\hbar c/k_B T$. For parallel plates, $F/A$ must be
proportional to $l_1 l_2$, and the problem reduces to the
determination of a function of $l_3$ and $l_T$. If one introduces
periodic orbits between the plates, then the appropriate surface
in $l_3$ and $l_T$ is the surface of a cylinder of length $l_3$
and circumference $l_T$. Periodic classical paths involve lengths
$2 n_3 l_3$, for $n_3$ to and fro motions between the plates, and
$n_T l_T$, for circling $n_T$ times around the cylinder. Since the
coordinate system is Cartesian, the length of the shortest orbit
that circles the cylinder $n_T$ times and goes back and forth
between the plates $n_3$ times is just\equ{length}. For a proper
derivation of the introduction of a fourth dimension, see\cite{m}
and Sec.~5.

\setcounter{equation}{0}
\section{AGREEMENT WITH A PREVIOUS RESULT}
We now show that $F^T/{\cal A}$ as given by\equ{forceT} agrees
with the force per unit area obtained previously\cite{l}. Using
Eqs.(3.8), (3.10), (3.11), (3.12), line~2 of (3.13), and (3.14)
of\cite{l}, we find
\be{forceSM}
\frac{F^T}{{\cal A}}=-\frac{1}{4\beta
l_3^3}\sum_{n=-\infty}^\infty c(n\hbar c/(2k_B T l_3))\ ,
\ee
where
\be{c}
c(\alpha) =-\frac{\beta}{4\pi^2\alpha
l_3}\frac{d^2}{d\alpha^2}J(\alpha)
\ee with
\be{J}
J(\alpha)=\int_0^\infty dy\frac{\sin(\alpha y)}{e^y-1}\ .
\ee
We obtain a different form for $c(\alpha)$ from that obtained
in\cite{l} by expanding the denominator of\equ{J},
\be{expand}
J(\alpha)={\rm Im}\,\sum_{m=1}^{\infty}\int_0^\infty dy
e^{(i\alpha-m) y}=\sum_{m=1}^{\infty}\frac{\alpha}{m^2+\alpha^2}\
,
\ee
in agreement with\equ{exp1}. Performing the differentiation leads
to
\be{cc}
c(\alpha) =\frac{\beta}{2\pi^2 l_3}\sum_{m=1}^{\infty}\frac{3
m^2-\alpha^2}{[m^2+\alpha^2]^3}\ .
\ee
Insertion of\equ{cc} with $\alpha=n \hbar c/(2 k_B T l_3)=n
l_T/(2 l_3)$ in\equ{forceSM} reproduces\equ{forceT}.

\setcounter{equation}{0}
\section{ELECTROMAGNETIC VERSUS SCALAR FIELDS}
We are primarily interested in electromagnetic fields and we
therefore comment on the relationship between the free energy
$A^T_{sc}$ of a scalar field and the free energy of an
electromagnetic field, denoted, as elsewhere, by $A^T$. The
boundary conditions for scalar and electromagnetic fields are
different, and further, for the latter case, both the condition
${\bf div~E}=0$, which has no scalar field analog, and
polarization must be accounted for; these three differences are
not, of course, unrelated. Electrodynamics assumes different forms
in spaces of a different number $D$ of spatial dimensions and we
will consider only $D=3$. We will further limit our considerations
to a rectangular box, with arbitrary dimensions $l_1$ by $l_2$ by
$l_3$.

As is well known, the condition ${\bf div~E}=0$ reduces the
number of independent components of $\bf E$ to at most two and for
a perfectly conducting box gives conditions on the quantum numbers
$n_1, n_2$ and $n_3$ of the modes. There are two independent
components if none of the $n_i$ vanish, one if one of the $n_i$
vanishes, and none if two of the $n_i$ vanish. In the periodic
orbital approach one considers classical ray's; in this case the
condition ${\bf div~E}=0$ implies certain phase changes of each of
the two independent polarizations upon reflection.

We consider polarization effects for three cases, i) $l_1\gg l_3$
and $l_2\gg l_3$, ii) $l_1\gg l_2$ and $l_1\gg l_3$, and iii)
arbitrary values of $l_1,l_2$ and $l_3$. For case i), the orbitals
pass back and forth between, and perpendicular to, the two $l_1$
by $l_2$ walls, the two states of polarization behave
identically, and the effective electromagnetic response function,
$\delta g_{el}$, is twice that of the relevant scalar response
function $\delta g_{sc}$. For case ii), the situation is the same
for orbitals which pass back and forth between the $l_1$ by $l_3$
or the $l_2$ by $l_3$ walls. For orbitals which reflect off four
walls, the pair of $l_1$ by $l_3$ and the pair of $l_2$ by $l_3$
walls, the polarization vector $\hat\epsilon$ is unchanged on
reflection for $\hat\epsilon$ perpendicular to the plane of
scattering; $\hat\epsilon$ is changed on reflection for
$\hat\epsilon$ in the plane of scattering, but returns to its
original value at the end of the (planar) periodic orbit, and
once again we have, effectively, $\delta  g_{el}=2 \delta
g_{sc}$. The situation is quite different for case iii), since
periodic orbits that reflect off all six walls are
{\it not} planar. [Periodic rays that reflect off all six walls
cannot be planar, since no plane can cut all surfaces of the
rectangular three dimensional volume.] To visualize one such
non-planar orbit consider a classical ray that is close to and
parallel to one of the spatial diagonals of the parallelepiped --
not contained in any of the surfaces -- and lies in one of the
diagonal planes that contains the diagonal and two of the corners
of the parallelepiped not connected by the diagonal. After
reflection off all three walls at one of the corners, the ray, for
symmetry reasons, is again parallel to the diagonal of the
parallelepiped and in the same plane as before reflection at the
corner. After reflecting off the other three walls at the
diagonally opposite corner, the path of the ray closes to a
periodic one. If the initial ray is chosen very close to the
diagonal, the length of this periodic orbit at the very least is
close to twice the length of the diagonal of the parallelepiped,
i.e.,  $L(1,1,1)=[(2l_1)^2+(2 l_2)^2+(2l_3)^2]^{1/2}$. The length
of the periodic classical ray that reflects once off each pair of
parallel walls in reality does not depend on how close to the
diagonal the incident ray is chosen and thus {\it is} twice the
length of the diagonal. The calculation would be much more
complicated if that were not the case. More complicated
non-planar periodic paths, that reflect a different number of
times off each pair of parallel surfaces are obtained by changing
the initial direction of the ray. For non-planar orbits there is
a mixing of the two states of polarization and $\delta g_{el}$
generally is not just $ 2 \delta g_{sc}$; the contribution to
$\delta g_{el}$ of non-planar orbits is proportional to the trace
of an orthogonal two-by-two polarization matrix. The existence of
non-planar periodic orbits for case~iii), and the consequent
non-diagonality of the polarization matrix, make less surprising
the fact that ${\cal E}_{\rm Cas}$ in case~iii) depends\cite{Luk}
not only on the 3-dimensional path lengths,
\be{3d}
L(n_1,n_2,n_3)=[(2l_1n_1)^2+(2l_2n_2)^2+(2l_3n_3)^2]^{1/2},
\ee
but also on the one-dimensional path lengths $L(n_i)=2l_i n_i$
for $i=1,\dots,3$. The presence of one-dimensional paths can also be 
associated with the restriction of the quantum numbers due to the 
boundary conditions. Thus for scalar fields satisfying Dirichlet 
boundary conditions in a rectangular cavity -- the eigenfunctions are 
products of sines -- none of the $n_i$ quantum numbers vanish.
Finally, we note that we stated in\cite{n} that
the semiclassical approach was exact if any of the ratios formed
from $l_1,l_2$ and $l_3$ was arbitrarily large, as is the case
for i) and ii) but not for iii). The restriction was necessary
because we had assumed that $\delta g_{el}=2\delta g_{sc}$.

We remark, parenthetically, that the four dimensional space
characterized by $l_1,l_2,l_3$ and $l_T$, with three spatial
dimensions and one periodic dimension (with a circumference
proportional to the inverse temperature), is not equivalent to
electrodynamics in four spatial dimensions;  ${\bf div~E}\neq 0$
for the latter space.


\begin{thebibliography}{99}
\bibitem{sumis} Ignoring corrections of order
$\epsilon$, Im$Q$ follows by evaluating the geometric series;
Re$Q$ follows by writing it as $\pi\sum_{-\infty}^\infty
\phi_m(y)\phi_m^*(-y)$, where $y=\beta x/2$ and
$\phi_m(y)=(2\pi)^{-1/2} \exp(imy)$, and using completeness of
these functions on the interval $(2n-1)\pi \leq y<(2n+1)\pi$ for
any $n$.
\end{thebibliography}
\end{document}